\documentclass[10pt,lettersize,journal]{IEEEtran}
\usepackage{amsmath,amssymb,amsfonts}
\usepackage{algorithmic}
\usepackage{algorithm}
\usepackage{array}
\usepackage[caption=false,font=normalsize,labelfont=sf,textfont=sf]{subfig}
\usepackage{textcomp}
\usepackage{stfloats}
\usepackage{url}
\usepackage{verbatim}
\usepackage{graphicx}
\usepackage{cite}
\usepackage{color}
\usepackage{flushend}
\begin{document}

\title{Jamming Suppression Via Resource Hopping in High-Mobility OTFS-SCMA Systems}

\author{Qinwen~Deng,~\IEEEmembership{Graduate Student Member,~IEEE, }Yao~Ge,~\IEEEmembership{Member,~IEEE, }and Zhi Ding,~\IEEEmembership{Fellow,~IEEE}
\thanks{This material is based upon work supported by the National Science Foundation under Grant No. 2009001 and No. 2029027.}
\thanks{Qinwen~Deng and Zhi~Ding are with Department of Electrical and Computer Engineering, University of California at Davis, Davis, CA 95616 USA (e-mail: mrdeng@ucdavis.edu; zding@ucdavis.edu).}
\thanks{Yao~Ge is with the Continental-NTU Corporate Lab, Nanyang Technological University, Singapore 639798 (e-mail: yao.ge@ntu.edu.sg).}}


\maketitle

\IEEEpubid{\begin{minipage}{\textwidth}\ \\[12pt] \centering
  \copyright 2023 IEEE. Personal use is permitted. Permission
from IEEE must be obtained for all other uses, in any current or future
media, including reprinting/republishing this material for advertising or
promotional purposes, creating new collective works, for resale or
redistribution to servers or lists, or reuse of any copyrighted
component of this work in other works.
\end{minipage}} 

\IEEEpubidadjcol

\begin{abstract}
This letter studies the mechanism of uplink multiple access and jamming suppression in an OTFS system. Specifically, we propose a novel resource hopping mechanism for orthogonal time frequency space (OTFS) systems with delay or Doppler partitioned sparse code multiple access (SCMA) to mitigate the effect of jamming in controlled multiuser uplink. We analyze the non-uniform impact of classic jamming signals such as narrowband interference (NBI) and periodic impulse noise (PIN) in delay-Doppler (DD) domain on OTFS systems. Leveraging turbo equalization, our proposed hopping method demonstrates consistent BER performance improvement under jamming over conventional OTFS-SCMA systems compared to static resource allocation schemes.

\end{abstract}


\section{Introduction}

\IEEEPARstart{H}{igh} mobility networking applications, 
such as hyper-high-speed railway (HSRs), unmanned aerial vehicles (UAVs), satellites, and vehicle-to-everything (V2X), represent some of the most important directions and exciting opportunities in future wireless communication networks\cite{b1,b2}. One of the most challenging environments is the doubly selective fading channels. Such commonly seen wireless channels under high mobility can lead to significant inter-carrier interference (ICI) in conventional orthogonal frequency division multiplexing (OFDM) systems and degrade the performance\cite{b5}.

Generally attributed to \cite{b6}, orthogonal time frequency space (OTFS) modulation has demonstrated strong error performance over conventional OFDM systems in doubly selective fading channels\cite{p1,p2}. By transforming the time-varying channel in time-frequency domain into a quasi-time-invariant channel in delay-Doppler (DD) domain and modulating the information symbols in DD domain, OTFS can effectively leverage diversity gain in both delay and Doppler channel domains, thereby achieving robust wireless connectivity even in the presence of significant channel Doppler spreads\cite{b7}. Previous works have addressed the bit error rate (BER) performance analysis and diversity benefits of OTFS through pairwise-error probability (PEP) in \cite{pep1,pep2,pep3} and signal-to-noise-plus-interference-ratio (SINR) in \cite{sinr1}.

To meet the growing needs of massive connectivity in future networks, non-orthogonal multiple access (NOMA) has also been jointly considered with OTFS. The authors in \cite{OTFS_NOMA1,OTFS_NOMA2} combined power-domain (PD) NOMA with OTFS and proposed a new protocol, where users with high mobility are grouped and served in the DD domain, and the users with low mobility are grouped and served in time-frequency (TF) domain. However, the use of unrealistic ideal bi-orthogonal OTFS pulses impedes the practical deployment 
of OTFS and OTFS-NOMA. An improved OTFS-based NOMA (OBNOMA) was introduced in \cite{b7}, in which practically implementable rectangular pulses were used with OTFS. Additionally, code-domain NOMA, such as sparse code multiple access (SCMA), was also introduced in OTFS system \cite{OTFS_SCMA}, where message passing (MP) algorithm was applied to detect the users' symbols. {As an important technology to fulfill the massive connectivity requirement in 5G and beyond networks \cite{OTFS_SCMA}, SCMA can suppress narrowband interference in the next generation Internet of things (IoT) \cite{FH_SCMA}. Therefore, we focus on an OTFS system with SCMA.}

The proposed integration of SCMA with OTFS \cite{OTFS_SCMA} requires that each user's signal be transmitted over a fixed DD domain regions for successive transmissions of OTFS blocks. As a result, BER performance of users under typical jamming or interference may suffer severely. To address such related issue in the conventional OFDM-SCMA system, the authors in \cite{FH_SCMA} relied on frequency hopping mechanism for the SCMA system. Furthermore, an inter-resource-blocks hopping SCMA system is proposed and analyzed in \cite{FH_SCMA2}, where physical resource-blocks (RBs) utilized by codewords are randomly hopped over the RB-group.

Inspired by recent advances, this work proposes a novel DD domain resource hopping method to mitigate jamming in multiuser OTFS-SCMA systems. We analyze the non-uniform impact of classic jamming signals, such as narrowband interference (NBI) and periodic impulse noise (PIN), on the DD domain by showing that the corresponding jamming power will focus on certain Doppler and delay bins of DD domain, respectively. To mitigate the negative impact of such jamming signals, we present customized resource hopping methods for multiuser OTFS-SCMA systems. Our simulation tests demonstrate consistent improvement of BER performance for our proposed resource hopping OTFS-SCMA under jamming over static resource allocation in uplink multiple access.
\vspace{-2mm}
\IEEEpubidadjcol

\section{Jamming Signal Analysis}
\label{sec:JammingAnalysis}
We first discuss and analyze the effect of classic jamming signals that may cause non-uniform power distributions on DD domain of OTFS system. For simplicity, the lattice in DD domain of OTFS system is denoted as
$$\Gamma \hskip -2pt =\hskip -2pt \left\{\hskip -2pt {\left(\frac{\ell}{{M\Delta f}},\frac{k}{{NT}}\right),\ell = 0, \cdots ,M - 1;k = 0, \cdots ,N - 1} \right\}$$
where $M$ and $N$ are the number of subcarriers and time slots; $T$ is the symbol period, which should be larger than the maximal channel delay spread; $\Delta f = \frac{1}{T}$ is the subcarrier spacing, which should be larger than the maximal Doppler spread. {Without loss of generality, the received time domain jamming signal $\mathbf{n}_J\in \mathbb{C}^{MN\times 1}$ will demodulate and transform to DD domain on the receiver of OTFS systems.} When using rectangular waveform in OTFS system, the affect of jamming signal vector $\mathbf{n}_J$ on DD domain is given by
\begin{equation}
\label{eq:1}
	\mathbf{w}_J = (\mathbf{F}_N\otimes\mathbf{I}_M)\mathbf{n}_J,
\end{equation}
where $\otimes$ denotes the Kronecker product. Let $\mathbf{W}_J\in\mathbb{C}^{M\times N}$ and $\mathbf{N}_J\in\mathbb{C}^{M\times N}$ be the matrix that is devectorized by $\mathbf{w}_J$ and $\mathbf{n}_J$, respectively. Using the property of Kronecker product, we can write (\ref{eq:1}) as
\begin{align} \label{eq:WN}
	\mathbf{W}_J = \mathbf{N}_J \mathbf{F}_N.
\end{align}
As the OTFS system can be implemented as a pre- and post-processing block to filtered OFDM system \cite{b6}, it is possible for classic jamming signals that are present in the OFDM system, such as NBI and PIN, to persist in OTFS systems. Without proper jamming suppression, the performance of the OTFS systems degrade significantly. To this end, we shall analyze the effects of NBI and PIN on OTFS systems before propose the efficient hopping scheme in the next section.

\subsection{Narrowband Jamming Signal} Without loss of generality, we consider only one narrowband jammer in our system. {According to \cite{nbi2}, the sampled narrowband jamming signal vector $\mathbf{n}_{NBI}\in \mathbb{C}^{MN\times 1}$ and the corresponding devectorized matrix $\mathbf{N}_{NBI} \in\mathbb{C}^{M\times N}$ }can be modeled as
\begin{align}
	\mathbf{n}_{NBI}[c] &= be^{j(2\pi\xi c \Delta f T/MN+\phi)}=be^{j(2\pi \xi 
 c/{MN}+\phi)},\\
    \mathbf{N}_{NBI}[\alpha,\beta] &= be^{j(2\pi \xi\alpha /MN+\phi)}e^{j(2\pi \xi\beta/N)},
\end{align}
{where $b$, $\xi\Delta f/N$ and $\phi$ are the amplitude, frequency and phase of the narrowband jamming signal, respectively; $\alpha\in [0,M-1]$ and $\beta\in [0,N-1]$ are the row and column indices of $\mathbf{N}_{NBI}$; $c=\alpha + \beta M$ is the index in $\mathbf{n}_{NBI}$. For the $\alpha$-th row of matrix $\mathbf{N}_{NBI}$, $be^{j(2\pi \xi\alpha /MN+\phi)}$ is a fixed complex value while the rest element $e^{j(2\pi \xi \beta/N)}$ is the phase rotation term. When $\xi$ has an integer value, the $N$-point row wise Fourier transform in (\ref{eq:WN}) yields the elements of matrix $\mathbf{W}_{NBI}$ as 
\begin{align*}
    \mathbf{W}_{NBI}[\alpha_W, \beta_W] = 
    \begin{cases}
    b\sqrt{N}e^{j(2\pi \xi\alpha_W /MN+\phi)},  & \text{if } \beta_{W} =\xi\\
    & \pmod{N}\\
    0, &\text{otherwise}
    \end{cases}
\end{align*}
i.e., the column vector of $\mathbf{W}_{NBI}$ has non-zero values only when $\beta_{W} = \xi \text{ (mod } N \text{)}$, as shown in Fig.~\ref{fig:jamming}(a).} In the multiuser OTFS system with fixed Doppler partition, such narrowband jamming signal is power efficient by decreasing the SINR of the jammed users, which leads to higher error rate.

\subsection{Periodic Impulse Noise} 
\begin{figure}
    \centering
    \subfloat[\label{fig:DelayJamming}]{
    \centering
    \includegraphics[width=0.48\linewidth]{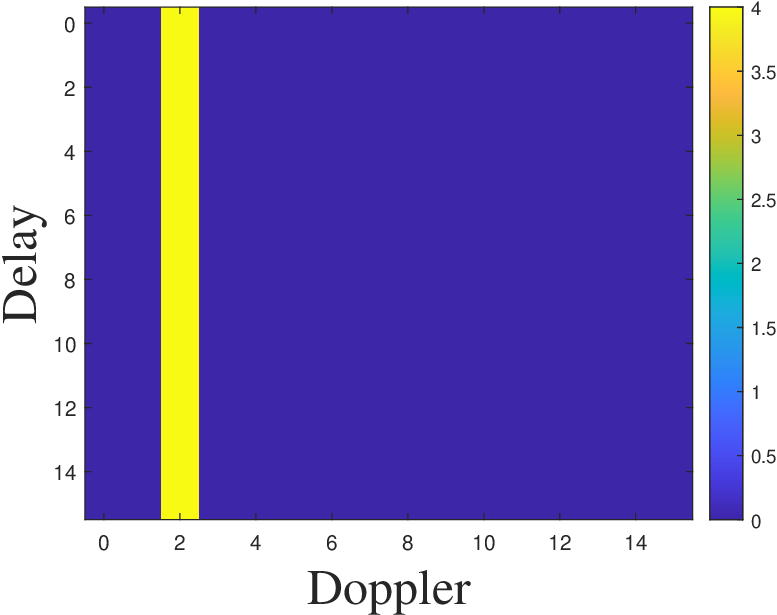}}
    \hfill
    \subfloat[\label{fig:DopplerJamming}]{
    \centering
    \includegraphics[width=0.48\linewidth]{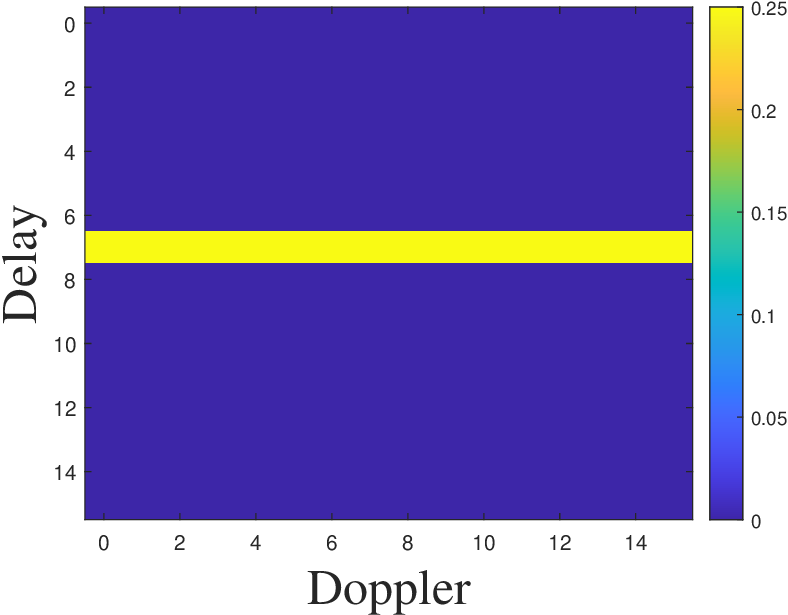}}
    \vspace{0mm}
    \caption{Impact of the NBI and PIN on DD domain in OTFS system with $M=N=16$. (a) Power distribution of the NBI on DD domain with the amplitude, frequency and phase of the NBI signal vector are $b=1$, $\xi=2$, $\phi=0$, respectively. (b) Power distribution of the PIN on DD domain with $\gamma_{PIN}=1$ and $T_p=NT$.}
    \vspace{0mm}
    \label{fig:jamming}
\end{figure}

PIN could also have non-uniform power distributions on the DD domain of an OTFS system. The sampled PIN vector $\mathbf{n}_{PIN}\in \mathbb{C}^{MN\times 1}$ and the corresponding matrix $\mathbf{N}_{PIN}\in \mathbb{C}^{M\times N}$ in one OTFS received block can be modeled as \cite{impulsen2}
\begin{align}
\mathbf{n}_{PIN}[c] &= 
\begin{cases}
    \gamma_{PIN},  & \text{if } c = c_{PIN}\\
    0, &\text{otherwise}
    \end{cases},\\
\mathbf{N}_{PIN}[\alpha,\beta] &= 
\begin{cases}
    \gamma_{PIN},  & \text{if } \alpha+\beta M = c_{PIN}\\
    0, &\text{otherwise}
    \end{cases},
\end{align}
where $\gamma_{PIN}\in \mathbb{C}$ is the constant complex response of the impulse. According to (\ref{eq:WN}), the element in the corresponding DD domain matrix $\mathbf{W}_{PIN}$ can be expressed as
\begin{align*}
    \mathbf{W}_{PIN}[\alpha_W, \beta_W] = 
    \begin{cases}
    \gamma_{PIN} e^{-j(2\pi \beta \beta_W/N)},  & \text{if } \alpha_W =\alpha \\
    0, &\text{otherwise}
    \end{cases},
\end{align*}
{where $\alpha = c_{PIN} \text{ (mod } M \text{)}$}. When the periodic time of the impulse is an integer multiple of $NT$, the power of all impulses in different OTFS transmission blocks will be located on the same $\alpha_W$-th row of $\mathbf{W}_{PIN}$, which corresponds to a fixed delay region, as shown in Fig.~\ref{fig:jamming}(b). Thus, such PIN is power efficient to jam a multiuser OTFS system with fixed delay partition.

\section{Proposed Hopping scheme}
\label{sec: system model}
In this section, we introduce the proposed resource hopping scheme for OTFS-SCMA systems to address the jamming signal effect on DD domain.

\vspace{0mm}
\subsection{Resource Hopping Scheme}
We consider a conventional multiuser OTFS system with one cyclic prefix (CP) and rectangular waveforms as in \cite{b3}. There are $U=GJ$ independent users in our OTFS system, partitioned into $G$ groups. For simplicity, we consider an uplink scenario and each UE group contains exactly $J$ users.

\begin{figure*}
    \centering
    \includegraphics[width=1.0\textwidth]{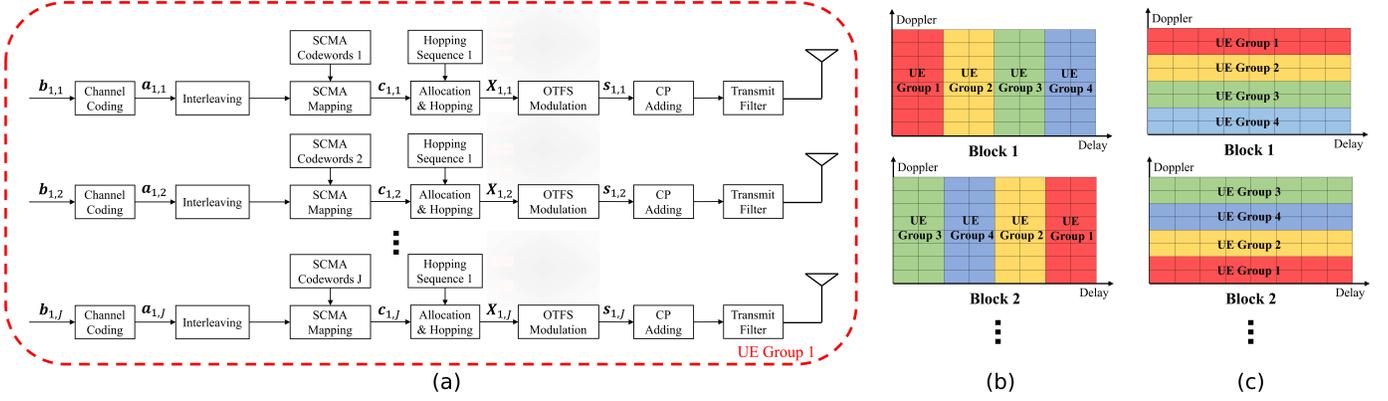}
    \vspace{-4mm}
    \caption{Block diagram of the transmitter of proposed hopping scheme for OTFS-SCMA systems. Different color regions of the DD domain OTFS transmission block correspond to different UE groups. All users in the same UE group will hop together in different OTFS transmission blocks. (a) Block diagram of all the transmitters of users in UE group 1. (b) OTFS transmission blocks in DD domain with resource hopping along delay partitions between 4 UE groups. (c) OTFS transmission blocks in DD domain with resource hopping along Doppler partitions between 4 UE groups.}
    \vspace{-2mm}
    \label{fig:SCMA}
    \vspace{-2mm}
\end{figure*}

{In this work, we allocate DD domain resources to different users so that all users in the same UE group utilize the same DD lattice region by SCMA, while different UE groups have non-overlapping DD lattice regions. Specifically, the DD lattice is partitioned by $G$ UE groups either along the delay axis or Doppler axis. Each UE group contains $J$ users that share transmission resource on DD domain by SCMA. The block diagram of the transmitter for the first UE group in the proposed OTFS hopping system is shown in Fig.~\ref{fig:SCMA}(a) as an example. For the $j$-th user in $g$-th UE group, its information bits $\mathbf{b}_{g,j}$ are first encoded by the channel coding such as low-density parity-check (LDPC) code. We denote the coded bits as $\mathbf{a}_{g,j}$. After interleaving, every $\log_2 Q$ coded bits are mapped into a complex symbol vector $\mathbf{c}\in \mathbb{A}_j$, where $\mathbb{A}_j$ denotes the SCMA codebook of the $j$-th user in the group with size $Q$.} Let $\mathbf{c}_{g,j}$ be the vector of all mapped symbol vectors of the $j$-th user in UE group $g$. All symbols in $\mathbf{c}_{g,j}$ are allocated to the partitioned DD lattice of the $g$-th UE group, which is selected by the random hopping sequence known at both transmitters and receiver. For delay partitioned OTFS system attacked by PIN, the proposed hopping scheme will change the allocated delay regions in different transmission blocks, as shown in Fig.~\ref{fig:SCMA}(b). {For example, as shown in Fig.~\ref{fig:SCMA}(b), the UE groups will be allocated along the delay domain by the order of [1,2,3,4] in the first transmission block, and in the next block this order will be changed to [3,4,2,1] by the hopping sequence.} Similarly, for Doppler partitioned OTFS system attacked by NBI, the proposed hopping scheme will change the allocated Doppler regions in different transmission blocks, as shown in Fig.~\ref{fig:SCMA}(c). After allocation, the DD domain signal $\mathbf{X}_{g,j}\in\mathbb{C}^{M\times N}$ is modulated by OTFS modulation with rectangular waveform to form a time domain signal $\mathbf{s}_{g,j}\in\mathbb{C}^{MN\times 1}$. A cyclic prefix (CP) is adding to the head of $\mathbf{s}_{g,j}$ to eliminate the interference between two successive OTFS transmission blocks.

In this work, we consider a time-varying multipath channel with channel delays and Dopplers that are not on the grid \cite{b7} for each user. The DD domain channel representation of $j$-th user in UE group $g$ can be written as
\begin{align}
	h_{g,j}(\tau,\nu) = \sum_{i=1}^{P_{g,j}}h_{g,j,i}\delta(\tau-\tau_{g,j,i})\delta(\nu-\nu_{g,j,i}),
\end{align}
where $P_{g,j}$ is the number of individual paths of $j$-th user in UE group $g$, $h_{g,j,i}$ is the channel gain of the $i$-th path of the user. $\tau_{g,j,i}$ and $\nu_{g,j,i}$ are the corresponding channel delay and Doppler shift, which can be further written as \cite{pep1}
\begin{align}
	\tau_{g,j,i}=\frac{\ell_{g,j,i}+\iota_{g,j,i}}{M\Delta f}, \nu_{g,j,i}=\frac{k_{g,j,i}+\kappa_{g,j,i}}{NT},
\end{align}
where $\ell_{g,j,i}$ and $k_{g,j,i}$ are the indices of channel delay and Doppler shifts, while $\iota_{g,j,i}\in (-1/2,1/2]$ and $\kappa_{g,j,i}\in (-1/2,1/2]$ are the fractional parts of the channel delay and Doppler shifts.

At the receiver side, the received time-domain signal denoted by $\mathbf{r}$ is the superposition of all user signals, channel noises, and jamming signals. After CP removal and OTFS demodulation, we transform the time-domain signal back to DD domain signal $\mathbf{Y}\in \mathbb{C}^{M\times N}$. 

Let $\mathbf{x}_{g,j}=\text{vec}(\mathbf{X}_{g,j})\in \mathbb{A}^{MN\times 1}$ and $\mathbf{y}=\text{vec}(\mathbf{Y})\in \mathbb{A}^{MN\times 1}$ be the vectorized form of the transmitted signal of $j$-th user in UE group $g$ and received signal $\mathbf{Y}$ in DD domain. Denote the equivalent $MN\times MN$ channel matrix from the $j$-th user in UE group $g$ to the receiver as
\begin{equation*}
	\mathbf{H}_{g,j}\hskip-2pt = \hskip-2pt\sum_{i=1}^{P_{g,j}}h_{g,j,i}(\mathbf{F}_N\otimes\mathbf{I}_M)\mathbf{\Pi}^{\ell_{g,j,i}+\iota_{g,j,i}}\mathbf{\Delta}^{k_{g,j,i}+\kappa_{g,j,i}}(\mathbf{F}_N^H\otimes\mathbf{I}_M),
\end{equation*}
and denote the full ${MN\times MNGJ}$
channel matrix from the users to the receiver as
\begin{equation}
\mathbf{H}=[\mathbf{H}_{1,1}^T, \mathbf{H}_{1,2}^T, \cdots, \mathbf{H}_{g,j}^T, \cdots,\mathbf{H}_{G,J}^T ]^T.
\end{equation}
Then the relationship between all $\mathbf{x}_{g,j}$ and $\mathbf{y}$ in the proposed hopping-based OTFS-SCMA system is given by
\begin{align}
	\mathbf{y} =\mathbf{H} \mathbf{x}+\mathbf{w},
    \label{y=Hx}
\end{align}
where $\mathbf{x} = [ \mathbf{x}_{1,1}^T, \mathbf{x}_{1,2}^T, \cdots, \mathbf{x}_{g,j}^T, \cdots,\mathbf{x}_{G,J}^T]^T$ represents transmitted users signals and 
\begin{align} \label{eq:n2}
	\mathbf{w}=(\mathbf{F}_N\otimes\mathbf{I}_M)(\mathbf{n}+\mathbf{n}_J)
\end{align}
is the vector superposition of the transformed noise and jamming signal on DD domain. Note that $\mathbf{\Pi}$ is the shifting matrix 
\begin{align}
	\mathbf{\Pi} = \begin{bmatrix}
		0 & \cdots & 0 & 1\\
		1 & \ddots & 0 & 0\\
		\vdots & \ddots & \ddots & \vdots\\
		0 & \cdots & 1 & 0
	\end{bmatrix};
\end{align}
$\mathbf{\Delta}$ is a diagonal matrix defined as $\mathbf{\Delta}=\text{diag}\{1,e^{\frac{j2\pi}{MN}},\ldots,e^{\frac{j2\pi(MN-1)}{MN}}\}$; $\mathbf{n}\in \mathbb{C}^{MN\times 1}$ and $\mathbf{n}_J\in \mathbb{C}^{MN\times 1}$ are the time domain noise vector and jamming signal vector by sampling the noise signal and jamming signal at the receiver with the sampling period $T_s=T/M$, respectively.
\vspace{0mm}

\subsection{Turbo Receiver Equalization}

\begin{figure}
    \centering
    \includegraphics[width=0.48\textwidth]{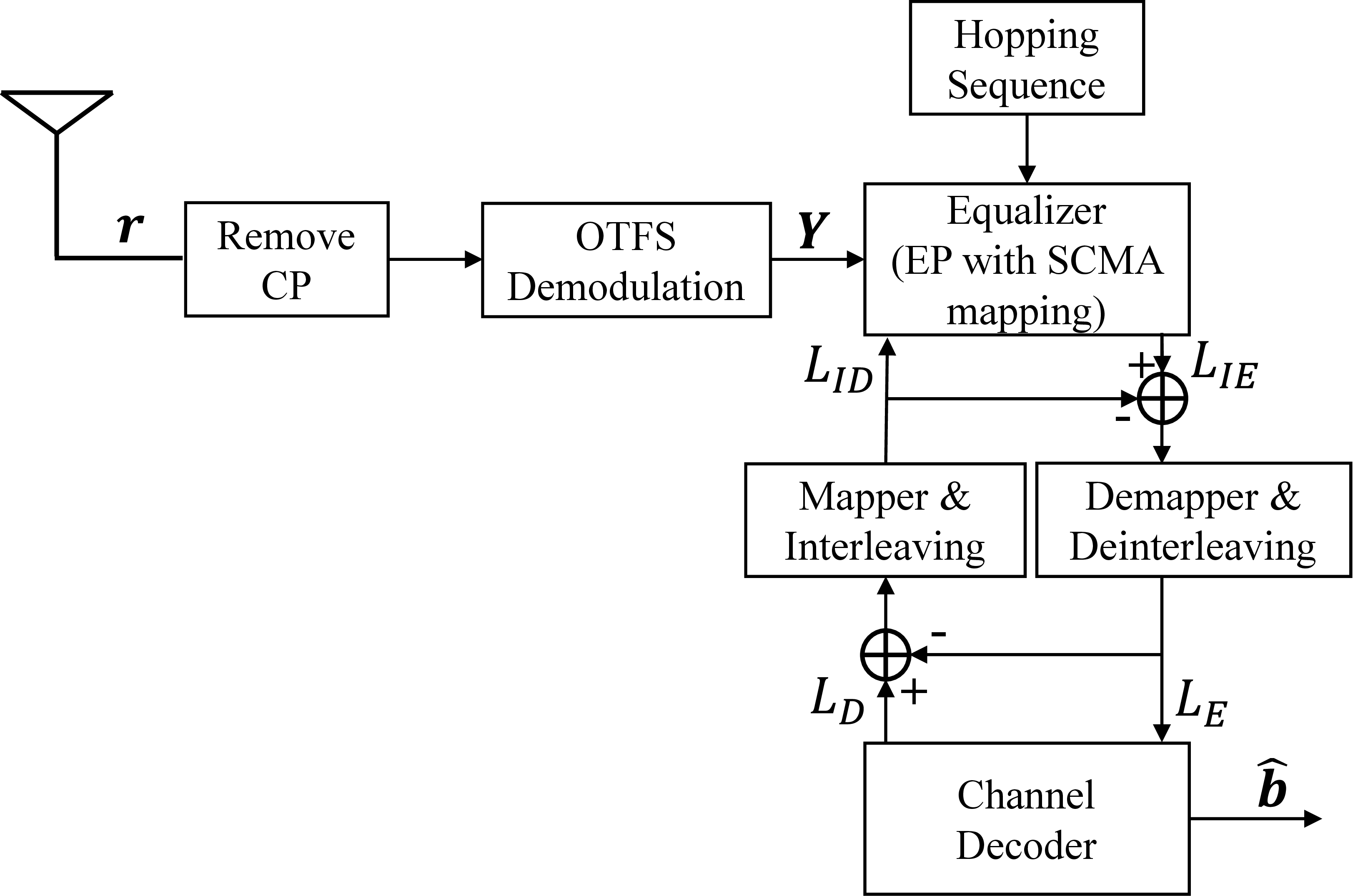}
    \vspace{0mm}
    \caption{Block diagram of the turbo receiver for proposed OTFS-SCMA hopping system.}
    \vspace{-4mm}
    \label{fig:receiver}
\end{figure}

{At the receiver, we develop a turbo structure of equalizer and decoder to recover the information bits of all users from received signal with jamming signal and channel noise, as shown in Fig.~\ref{fig:receiver}. We assume that all effective channel response matrices $\mathbf{H}_{g,j}$, the hopping sequence and interleaving seeds are known to the receiver. By applying the Gaussian approximation with expectation propagation (GAEP) detector to the vectorized DD-domain received signal $\mathbf{y}$ and overall equivalent channel matrix $\mathbf{H}$ from (\ref{y=Hx}), we obtain the probability matrix for $\mathbf{x}$, as described in \cite{OTFS_alg}. Then, we convert the probability matrix to the log-likelihood matrix $L_{IE}$ for $\mathbf{x}$ as the output of the equalizer. The extrinsic log-likelihood matrix is calculated by subtracting feedback log-likelihood matrix $L_{ID}$ from the decoder. This extrinsic log-likelihood matrix is convert to the log-likelihood matrices of all $\mathbf{a}_{g,j}$, denoted as $L_E$, by demapper and deinterleaving, and fed to the channel decoder as the input. Similarly, the feedback extrinsic log-likelihood matrix is calculated by subtracting $L_E$ from the output log-likelihood matrix of the channel decoder, denoted as $L_D$. The mapper and interleaving block will transform this feedback extrinsic log-likelihood matrix to $L_{ID}$, which is the feedback log-likelihood matrix of $\mathbf{x}$. Finally, $L_{ID}$ is sent to the equalizer as the \textit{a priori} information and completes the loop. The receiver terminates after $\mathcal{L}$ loops, and finally outputs the estimation of all information bits $\widehat{\mathbf{b}}$.} {Suppose the total number of loop is $n_l$, and in each loop the GAEP detector in the equalizer has $n_c$ iterations. Then the complexity of the GAEP detector equals $n_l n_c \mathcal{O}(6\overline{S}D+\overline{S}DQ+\frac{MNJDQ}{K})$ by \cite{OTFS_alg}, where $\overline{S}$ is the total number of nonzero elements in $\mathbf{H}$ whereas $K$ is the dimension of each SCMA codeword, and $D$ is the number of non-zero entries in each SCMA codeword.}

 \section{Numerical result}

In this section, we describe the setup of our tests and present the error performance of the proposed hopping OTFS-SCMA system. We consider a multiuser OTFS system with number of users $U=24$. We partition the 24 users into $G=4$ groups, with each group containing $J=6$ users. We set carrier frequency to 4 GHz and subcarrier spacing $\Delta f=15$ kHz. We consider a DD lattice with $M=128$ and $N=16$. Without loss of generality, we partition the resources on DD domain equally into $G$ parts either along delay axis or Doppler axis, as mentioned in Section~\ref{sec: system model}. We use a (3, 6)-regular LDPC of length 256 with a rate of 1/2 based on the progressive-edge growth (PEG) algorithm \cite{LDPC_codes} and apply belief propagation \cite{LDPC_alg} with a maximum number of 10 iterations as channel decoder. We use the SCMA codebooks provided by \cite{SCMA_codes} for $Q=4$ with $J=6$ users spreading over shared $K=4$ resources, having $D=2$ non-zero entries in each codeword. We generate the hopping sequences of length $G$ randomly such that each user has the same opportunity to be allocated to any partitioned parts. The transmission power of each user is the same.

 \begin{figure}[t]
    \centering
    \subfloat[\label{fig:Delay}]{
    \centering
    \includegraphics[width=0.66\linewidth]{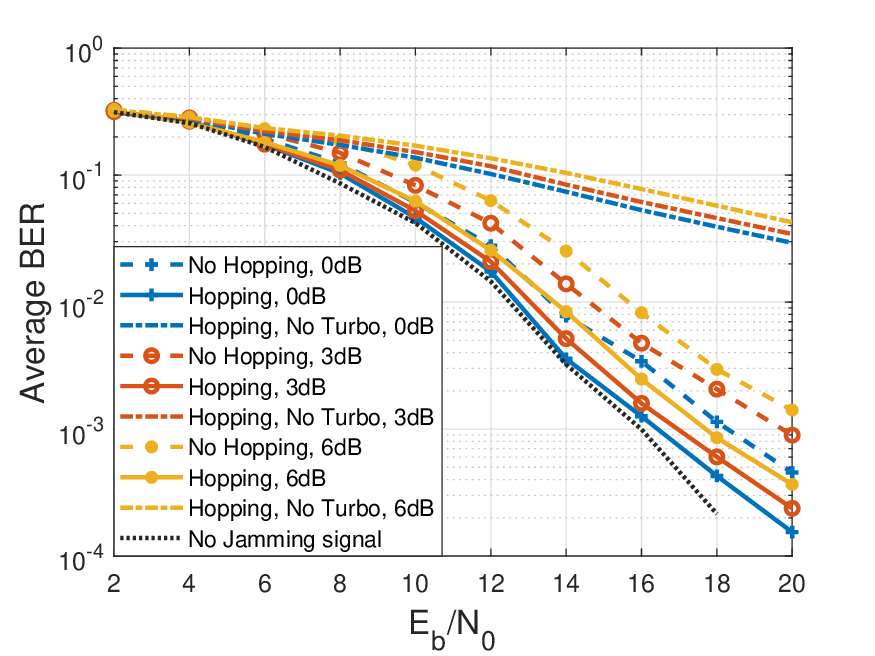}}
    \\
    \vspace{-2mm}
    \subfloat[\label{fig:Doppler}]{
    \centering
    \includegraphics[width=0.66\linewidth]{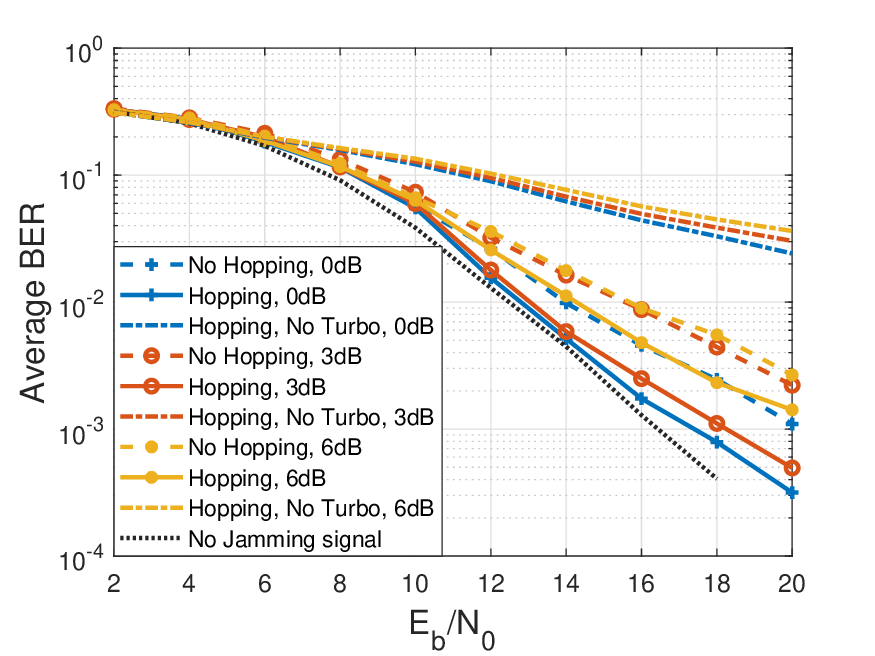}}
    \vspace{0mm}
    \caption{Average BER of resource hopping for both delay and Doppler partitioned OTFS-SCMA systems with different jamming signal-to-noise ratios.  (a) average BER of resource hopping for delay partitioned OTFS-SCMA system with 3 PIN jamming signals of different $T_i$. (b) average BER of resource hopping for Doppler partitioned OTFS-SCMA system with 3 NBI jamming signals of the same power.}
    \label{fig:Delay1}
    \vspace{-2mm}
\end{figure}

We set the parameters of channels in our experiment as follows. We let the number of individual paths $P_{g,j}=5$
be the same for all users. We consider channel gains $h_{g,j,i}$ as independent and identically distributed (i.i.d.) Gaussian random variables with zero mean and variance $1/(2UP_{g,j})$ for real and imaginary part. 
We also let the maximum channel delay $\tau_{max}=2/(M\Delta f)$ and the channel delays for all paths of all users be i.i.d. variables from uniform distribution between $0$ and $\tau_{max}$. We set the velocity of all mobile users
to $120$ km/h, leading to a maximum Doppler shift $\nu_{max}=444.44$ Hz. We simulate
the Doppler shift for each path of each user by using the Jakes formulation \cite{Jake}, i.e., $\nu_{g,j,i}=\nu_{max}\cos(\theta_{g,j,i})$, where all $\theta_{g,j,i}$s are i.i.d. variables from uniform distribution between $-\pi$ and $\pi$. From the analysis of Section~\ref{sec:JammingAnalysis}, PIN is more efficient to jam the delay partitioned OTFS system, while NBI is more efficient to jam the Doppler partitioned OTFS system. Thus, we analyze our proposed resource hopping scheme in two scenarios: PIN jammed delay partitioned OTFS system and NBI jammed Doppler partitioned OTFS system. We configure the PIN parameters 
such that $T_p=NT$ and randomly select the initial time interval $T_i$ from
the set that can jam the target users. We set the parameters of NBI 
such that $\phi$ is uniform between $[-\pi,\ \pi]$ and 
$\xi\in [0,M-1]$ is randomly selected in the set that can jam the target users. We let
one UE group undergo jamming and regard all users within this group 
as jammed users. We set the power of all jamming signals in both hopping scenarios 
to be the same, while setting the total jamming power 
according to the jamming signal-to-noise ratio.

Fig.~\ref{fig:Delay1} illustrates the performance of the proposed hopping schemes for OTFS-SCMA system.
We tested three jamming signal-to-noise ratios at
0dB, 3dB, and 6dB in both hopping scenarios. Without changing
the power of each jamming signal, 
we adjust the noise power and transmit signal power 
according to the $E_{b}/N_{0}$. The results in Fig.~\ref{fig:Delay1}(a) and Fig.~\ref{fig:Delay1}(b) clearly show that under the proposed resource hopping scheme,
the average BER of users under jamming is far superior to those without resource hopping.
The benefit is evident in both delay and Doppler partitioned OTFS-SCMA systems.
These results and analyses confirm that our proposed resource hopping scheme 
can efficiently mitigate the jamming effect of NBI and PIN in OTFS-SCMA systems.
\vspace{0mm}
\section{Conclusion}
This work investigates the non-uniform effect of jamming signals on DD domain for high mobility users in OTFS systems. To mitigate the effect of jammers or interferences on multiuser access in OTFS, we develop a simple but effective resource hopping method in OTFS-SCMA system. By hopping along delay or Doppler axis for different OTFS blocks, multiple access users can effectively suppress severe interference caused by common jamming signals such as PIN and NBI. Our simulation results demonstrate that the proposed hopping scheme can efficiently improve the average BER performance for multiple access users under jamming in OTFS-SCMA systems.



\end{document}